\documentclass[letterpaper, 10 pt, conference]{ieeeconf}
\IEEEoverridecommandlockouts
\usepackage{times}
\usepackage{fancyhdr}
\usepackage{float}
\usepackage{listings}
\usepackage{booktabs}
\usepackage{subfig, bm}

\usepackage{amsthm}
\usepackage{tikz}
\usepackage{amssymb}
\usepackage{mdwlist}
\usepackage{hyperref}
\hypersetup{colorlinks,linkcolor={blue},citecolor={blue}} 
\usepackage{amsmath,graphicx,xcolor}
\usepackage{subfiles}
\usepackage{import}
\usepackage{dsfont}
\usepackage[font=small,labelfont=bf]{caption}
\usepackage{mathtools}
\usepackage{algorithm}
\usepackage{algorithmic}
\usepackage{sjmacros}
\usepackage{thmstyles}
\usepackage{stfloats}
\usepackage{cuted}
\title{\LARGE \bf
OUTformation: Distributed Data-Gathering with Feedback under Unknown Environment and Communication Delay Constraints
}

\author{
SooJean Han$^{1*}$, Michelle Effros$^{1}$, Richard M. Murray$^{1}$
\thanks{$^{1}$SooJean Han, Michelle Effros, and Richard M. Murray are with the Division of Engineering and Applied Science at California Institute of Technology, Pasadena CA.}
\thanks{$^{*}$Corresponding author. Email: {\tt\small soojean@caltech.edu}}
\thanks{This paper is based on work supported by the National Science Foundation Graduate Research Fellowship under Grant No. DGE‐1745301.}
}

\begin{document}
\maketitle
\thispagestyle{plain}
\pagestyle{plain}
\allowdisplaybreaks

\begin{abstract}
    Towards the informed design of large-scale distributed data-gathering architectures under real-world assumptions such as nonzero communication delays and unknown environment dynamics, this paper considers the effects of allowing feedback communication from the central processor to external sensors.
Using simple but representative state-estimation examples, we investigate fundamental tradeoffs between the mean-squared error (MSE) of the central processor's estimate of the environment state, and the total power expenditure per sensor under more conventional architectures without feedback (\textit{INformation}) versus those with broadcast feedback (\textit{OUTformation}).
The primary advantage of enabling feedback is that each sensor's understanding of the central processor's estimate improves, which enables each sensor to determine when and what parts of its current observations to transmit.
We use theory to demonstrate conditions in which OUTformation maintains the same MSE as INformation with less power expended on average, and conditions in which OUTformation obtains less MSE than INformation at additional power cost.
These performance tradeoffs are also considered under settings where environments undergo less variation, and sensors implement random backoff times to prevent transmission collisions.
Our results are supported via numerical studies, which show that the properties derived in theory still hold even when some of the simplifying assumptions are removed.
% 197 words.
% ; this allows us to discuss the implications for more complex settings.
% We employ two performance metrics, the power cost per sensor and the mean-squared error (MSE) of the central processor's estimate of the environment's state, to compare the two architectures; we demonstrate that , .
% We use simulations to demonstrate more complex tradeoffs such as the speed of change in the true environment state...
% feedback architectures and investigate their performance in distributed state estimation problems when there is limited information about the statistics of the environment and communication delays among the processors and sensors.
% We then use theoretical analysis and simulations to compare state estimation performance with and without feedback under simple settings.
\end{abstract}
\vskip.1cm

\newcommand{\sensor}{verification}
\newcommand{\trigger}{transmission}
\newcommand{\cpu}{fusion}

\makeatletter
\newcommand{\distas}[1]{\mathbin{\overset{#1}{\kern\z@\sim}}}%
\newsavebox{\mybox}\newsavebox{\mysim}
\newcommand{\distras}[1]{%
  \savebox{\mybox}{\hbox{\kern3pt$\scriptstyle#1$\kern3pt}}%
  \savebox{\mysim}{\hbox{$\sim$}}%
  \mathbin{\overset{#1}{\kern\z@\resizebox{\wd\mybox}{\ht\mysim}{$\sim$}}}%
}
\makeatother

%%%%%%%%%%%%%%%%%%%%%%%%%%%%%%%%%%%%%%%%%%%%%%%%%%%%%%%%%%%%%%
% INTRODUCTION
%%%%%%%%%%%%%%%%%%%%%%%%%%%%%%%%%%%%%%%%%%%%%%%%%%%%%%%%%%%%%%

\section{Introduction}
In previous literature, a variety of architectures for distributed data-gathering in large-scale cyberphysical network applications has been proposed with the aim of reducing unnecessary transmissions of redundant information and excess computation; see~\cite{castanedo13,liggins97} for surveys.
However, many methods consider only structural variations in the communication among lower-level nodes such as external sensors (e.g., hierarchical clustering~\cite{bandyopadhyay03,younis04}), and have rarely challenged the imposition that data flows only in a single direction from sensors on the network's edge to the higher-level processor at the network's center.
Some architectures allowing \textit{feedback communication} in the reverse direction, from the central processor to the external sensors, are developed under a diversity of simplifying assumptions.
In some feedback schemes for redundant transmission reduction (e.g.~\cite{chen15}), communication delays are omitted; this disregards issues pertaining to outdated data, which is important to distributed state estimation in real-world large-scale networks.
In others (e.g.~\cite{ge14outformation}), communication delays are included, but the approach for redundant transmission reduction assumes full knowledge of the environment dynamics.

While in practice, the best choice of architecture to use for distributed data-gathering varies by setting, imposing simplifying but impractical assumptions makes it difficult to properly motivate feedback, and make fair comparisons between architectures which employ feedback and those which do not.
Feedback can provide at least the following two advantages in large-scale distributed systems.
% BAD POINT 1: LARGE DELAYS.
First, by nature of the distributed architecture, data gathering and data processing are sequential actions; having a larger delay between the sensor transmission times and central processor receipt times can potentially lead to large estimation errors.
% BAD POINT 3: WHEN THERE IS A LARGE NUMBER OF SENSORS, REDUNDANT INFORMATION CAN BE ESPECIALLY COSTLY.
Second, for large-scale systems, a large number of sensors may be needed to survey the environment.
But independent sensors surveying a shared environment are likely to collect data which exhibits redundancies, and transmitting redundant information to the central processor is inefficient.

With these motivations in mind, the contributions of our paper are as follows.
We employ a modular framework for distributed state estimation under two important real-world constraints: nonzero communication delays and limited knowledge of the environment dynamics.
Because the type of feedback we consider broadcasts data ``out'' from the central processor to the sensors, we refer to this architecture as the \textit{OUTformation architecture} for data-gathering, as opposed to the \textit{INformation architecture} where all information flows from the sensors ``in'' to the central processor.
We use a modular framework representation of the INformation and OUTformation architectures to take a first step towards characterizing their relative performances with respect to two specific metrics: 1) the mean-squared error (MSE) of the central processor's estimate of the environment state, and 2) the total power expenditure of each sensor.
The importance of such a characterization arises from allowing users to determine which type of architecture is more suitable for use based on hardware specifications and limited knowledge of the environment, as well as providing essential insights to make parameter design choices for optimal performance in both MSE and power expenditure.

We theoretically analyze simple but insightful case studies demonstrating fundamental tradeoffs between the two performance metrics, and consider how the tradeoffs vary under different environment statistics and parameter design choices.
In particular, we show that the main advantage of enabling feedback is improving each sensor's estimate of the central processor's estimate, which allows each sensor to make more informed decisions about when and what to transmit, potentially reducing both 1) its uplink power expenditure, and 2) the MSE by minimizing the delays after which the central processor receives its transmissions.
We employ numerical simulations to break some of the simplifying assumptions made in our theoretical analysis, and show that the properties derived in theory still hold, which suggests that the implications of OUTformation architectures extend beyond the framework treated rigorously by the theory.

%%%%%%%%%%%%%%%%%%%%%%%%%%%%%%%%%%%%%%%%%%%%%%%%%%%%%%%%%%%%%%
% SETUP: DYNAMICS, SENSORS, PROBLEM GOAL
% MODULAR FRAMEWORK
%%%%%%%%%%%%%%%%%%%%%%%%%%%%%%%%%%%%%%%%%%%%%%%%%%%%%%%%%%%%%%

\section{INformation and OUTformation}\label{sec:setup}
\subsection{Problem Setup}\label{subsec:problem_setup}
We consider environments which evolve as a random walk given by
\begin{align}\label{eq:random_walk}
    \xvect((c+1)\Delta t) = \xvect(c\Delta t) + \sum\limits_{i=1}^n \delta_i(c)\dvect(c)
\end{align}
Here, $c\in\Nbb$, and $\Delta t{\,\in\,}\Rbb^{+}$ is the environment's transition period.
For each $i{\,\in\,}\{1,\cdots, n\}$ and $c\in\Nbb$, random variable $\delta_i(c)$ takes value $1$ with probability $p/2$, $-1$ with probability $p/2$, and $0$ with probability $1-p$ for some $p{\,\in\,}(0,1)$, and $\dvect(c){\,\in\,}\Rbb^n$ such that $d_i(c){\,\sim\,}\Ucal[\underline{d}, \overline{d}]$ is a uniformly-distributed random variable stepsize between constant values $0<\underline{d}<\overline{d}$.
% ; we refer to the interval of time $[c\Delta t, (c+1)\Delta t)$ as the \textit{environment interval $c$}.
% \revised{
\begin{definition}[Unknown Environment]\label{def:unknown}
    The titular ``unknown environment'' refers to the fact that the parameters $\Delta t, p, \underline{d}, \overline{d}$ are unknown to the central processor and all the sensors.
\end{definition}

We focus on architectures where a single central processor communicates with $M{\,\in\,}\Nbb$ independent external sensors.
Each sensor $j\in\{1,\cdots,M\}$ employs a linear measurement equation perturbed by additive, white Gaussian noise $\wvect_j(t){\,\sim\,}\Ncal(\textbf{0}, \sigma^2I_n)$ where $I_n{\,\in\,}\Rbb^{n\times n}$ is identity:
\begin{align}\label{eq:sensor}
    \yvect_j(a\tau_j) = C_j \xvect(a\tau_j) + \wvect_j(a\tau_j)
\end{align}
Here, $C_j{\,\in\,}\{0,1\}^{m_j \times n}$, $m_j\in\Nbb, m_j{\,<\,}n$, such that each row contains exactly one 1.
% and each column contains at most one 1. 
Each sensor $j$ makes one observation of the environment with \textit{sampling period} $\tau_j$ timesteps, and $a\in\Nbb$ is any number.
% For clarity of expression, we use the terminologies ``uplink'' and ``downlink'' communication in place of ``feedforward'' and ``feedback'' communication respectively.
% , and we denote the time series of vectors $\{\yvect(t_1), \cdots, \yvect(t_2)\} \triangleq \yvect(t_1:t_2)$ for any $0 \leq t_1 < t_2$.

Over some experiment duration $[0,T_{\text{sim}}), T_{\text{sim}}\in\Rbb^{+}$, the central processor's objective is to use the transmissions of each sensor to construct an estimate $\hat{\xvect}(t)$ of the environment state $\xvect(t)$.
Each sensor $j$'s objective is to determine when and which components $y_{jk}(t), k{\,\in\,}\{1,\cdots, m_j\}$ of $\yvect_j(t)$ it should transmit; we show that this is contingent upon how accurately the sensor can track the central processor's estimate $\hat{\xvect}(t)$.
% For concreteness, we perform the majority of our analyses using $M=2$ sensors, though we also discuss the implications of our findings to a general number $M\geq 2$ of sensors.
%

\begin{assumption}[Framework Assumptions]\label{assum:paper_assumptions}
    Each sensor $j$ operates on limited power sources (e.g. batteries).
    Communications in both uplink and downlink channels are noiseless and able to precisely encode/decode real numbers.
    To prevent potential collisions of transmissions, each sensor $j$ schedules \textit{random backoff times} $B_{jk}(t){\,\distas{d}\,}B$ for some random variable $B{\,\in\,}[0,\infty)$ distributed according to the cdf $F_B(s){\,\triangleq\,}\Pbb(B{\,\leq\,}s)$ for each component $k{\,\in\,}\{1,\cdots, m_j\}$ and time $t$.
    The $B_{jk}(t)$ are distributed independently of each other and of the noise process $\wvect_j(t)$.
    The central processor performs \textit{fusion} by averaging over the most recent sensor observations it received to construct an estimate $\hat{x}_i(t)$ for each component $i\in\{1,\cdots, n\}$.
\end{assumption}

\begin{remark}
    We emphasize that our goal is not to model any particular distributed data-gathering architecture in fine detail, but to provide insights into the distinctions of architectures with and without feedback from the central processor to the sensors.
    Thus, while more complex functionalities (e.g., optimized random backoff strategies~\cite{nutt82}, consensus-based fusion~\cite{olfati05}, change-detection~\cite{fearnhead19}) may be used than those described in~\assum{paper_assumptions}, we focus specifically on functionalities which need to be implemented differently based on whether an architecture has enabled feedback or not.
\end{remark}

\subsection{Modular Framework}\label{subsec:modular}
Under~\assum{paper_assumptions}, we distinguish between traditional INformation and OUTformation architectures in the following way.
The presence or absence of broadcast feedback (which pushes data ``outward'' from the central processors to the sensors) changes how accurately each sensor can track the central processor's estimate $\hat{\xvect}(t)$.

\begin{figure}
    \begin{center}
       \includegraphics[width=0.65\columnwidth]{../figures/diagram}
    \end{center}
    \caption{A modular representation of an OUTformation architecture for distributed state estimation under unknown environments and communication delay constraints, using a single central processor and a network of $M$ external sensors. 
    For each sensor, the \sensor{} rule (\defin{\sensor}) is highlighted in green, and the \trigger{} rule (\defin{\trigger}) is in blue.
    The broadcast rule (\defin{broadcast}) is in red; INformation architectures are depicted by removing the red broadcast arrows.
    }
    \label{fig:complete_model}
\end{figure}

\begin{definition}[INformation and OUTformation]\label{def:architecture}
    Under an \textit{INformation architecture}, each sensor's estimate of $\hat{x}_i(t), i\in\{1,\cdots, n\}$ is its own previous transmission to the central processor.
    Under an \textit{OUTformation architecture}, each sensor's estimate of $\hat{x}_i(t)$ is the more recent of either its own previous transmission or the latest broadcast received.
    A modular representation of the distributed data-gathering OUTformation architecture, which we use throughout the paper, is illustrated in~\fig{complete_model}.
\end{definition}

\begin{notation}\label{nota:superscript}
    Parameters are assigned the superscript ${(I)}$ for INformation architectures, and $(O)$ for OUTformation architectures.
    Let $\chi$ be a placeholder variable such that $\chi{\,=\,}I$ for an INformation architecture, and $\chi{\,=\,}O$ for OUTformation.
    We henceforth let $\hat{\xvect}^{(\chi)}{\,\in\,}\Rbb^n$ be the central processor's estimate under architecture $\chi\in\{I,O\}$.
\end{notation}

\begin{notation}\label{nota:full_vector_component}
    For each sensor $j{\,\in\,}\{1,\cdots, M\}$ and component $k{\,\in\,}\{1, \cdots, m_j\}$, we let $i_k{\,\in\,}\{1,\cdots, n\}$ correspond to the ``full state vector component'' from which observation $y_{jk}(t)$ is made, i.e. $x_{i_k}(t){\,\triangleq\,}C_{j,(k,\cdot)}\xvect(t)$, where $C_{j,(k,\cdot)}$ denotes the $k$th row of $C_j$ from~\eqn{sensor}.
\end{notation}

\begin{definition}[Shared vs. Unshared Components]\label{def:shared}
    For each sensor $j\in\{1,\cdots, M\}$, let $k{\,\in\,}\{1, \cdots, m_j\}$ be a component and $i_k\in\{1,\cdots, n\}$ be the corresponding full-vector component (see~\nota{full_vector_component}). 
    Then $k$ is said to be a \textit{shared} component if $x_{i_k}(t)$ is sensed by other sensors, and an \textit{unshared} component if $x_{i_k}(t)$ is sensed by only $j$.
    Define the set $\Ucal_j{\,\subseteq\,}\{1, \cdots, m_j\}$ to be the set of sensor $j$'s unshared components, and $\Scal_j\triangleq\Ucal_j^{c}$ to be the set of sensor $j$'s shared components.
\end{definition}

\begin{definition}[Broadcast Rule]\label{def:broadcast}
    OUTformation architectures employ the following \textit{broadcast rule} (red in~\fig{complete_model}): a component $i\in\{1,\cdots, n\}$ is scheduled to be broadcast if $\hat{x}_i^{(O)}(t) \neq \hat{x}_i^{(O)}(s)$, where $s<t$ is the time of the previous broadcast.
    All scheduled values $\hat{x}_i^{(O)}(t)$ are collected to form the dimension-reduced \textit{broadcast vector} $\hat{\xvect}^{(b)}(t)\in\cup_{i=1}^n\Rbb^i$ and sent to the sensors at once.
\end{definition}

% \revised{
\begin{assumption}[Components Being Broadcast]
    Broadcasts are made only for components which are shared in the sense of~\defin{shared}.
    % This is because in an environment which is unknown in the sense of~\defin{unknown}, there is no benefit to sensors receiving updates for components they do not observe.
\end{assumption}
% }

\begin{definition}[Verification Rule]\label{def:\sensor}
    For $\chi{\,\in\,}\{I,O\}$ as in~\nota{superscript} and each sensor $j{\,\in\,}\{1,\cdots, M\}$, we define $T_j^{(\chi)}$, a multiple of the sampling period $\tau_j$ from~\eqn{sensor}, to be the \textit{\sensor{} period} under architecture $\chi$.
    At time $t{\,\triangleq\,} aT_j^{(\chi)}$, $a\in\Nbb$, the \textit{\sensor{} rule} schedules the transmission of $y_{jk}(t)$, $k{\,\in\,}\{1,\cdots, m_j\}$ for time $t{\,+\,}B_{jk}(t)$ if $\beta_{jk}^{(\chi)}(t) = 1$, where $B_{jk}(t)$ is the random backoff time from~\assum{paper_assumptions}.
    Under an INformation architecture, we use:
    \begin{align}\label{eq:in_\sensor_rule}
        &\beta_{jk}^{(I)}(t) = \mathds{1}\left\{\abs{y_{jk}(s_{jk}) - y_{jk}(t)} \geq \epsilon\right\}
    \end{align}
    while under an OUTformation architecture, we use:
    \begin{align}\label{eq:out_\sensor_rule}
        &\beta_{jk}^{(O)}(t) = \begin{cases}
            \mathds{1}\{ \abs{\hat{x}^{(b)}_{i_k}(t) - y_{jk}(t)} \geq \epsilon \} &\text{ if } s^{*} \geq s_{jk}\\
            \beta_{jk}^{(I)}(t) &\text{ else}
        \end{cases}
    \end{align}
    Here, $s_{jk}{\,<\,}t$ is the time of sensor $j$'s previous transmission of component $k$, and $s^{*}{\,<\,}t$ is the time the broadcast value $\hat{x}^{(b)}_{i_k}(t)$ was created.
    Threshold parameter $\epsilon{\,>\,}0$ is chosen by design.
    The special case of \textit{event-triggered verification} occurs when the \sensor{} rule is checked with every observation sensor $j$ generates, i.e. $T_j^{(I)}{\,=\,}T_j^{(O)}{\,=\,}\tau_j$.
\end{definition}

\begin{definition}[Transmission Rules]\label{def:\trigger}
    Let $a{\,\in\,}\Nbb$, and let $\chi{\,\in\,}\{I,O\}$ be the placeholder variable defined in~\nota{superscript}.
    For each sensor $j$, let $T_j^{(\chi)}$ be the \sensor{} period from~\defin{\sensor}.
    At time $t{\,+\,}B_{jk}(t)$ where $t{\,\triangleq\,} aT_j^{(\chi)}$ and $B_{jk}(t)$ is the random backoff time from~\assum{paper_assumptions}, the \textit{\trigger{} rule} transmits $y_{jk}(t)$ for every component $k{\,\in\,}\{1,\cdots, m_j\}$, if $\beta_{jk}^{(\chi)}(t){\,=\,}1$ and $\gamma_{jk}^{(\chi)}(t,B_{jk}(t)){\,=\,}1$.
    Essentially, the transmission rule is a re-checking of the verification rule in case the sensor receives more up-to-date data about the central processor's estimate.
    Because each sensor only has their own local observations under an INformation architecture, we have:
    \begin{align}\label{eq:in_\trigger_rule}
        \gamma_{jk}^{(I)}(t,B_{jk}(t)) &= \beta_{jk}^{(I)}(t)
        % \mathds{1}\{\beta_{jk}^{(I)}(t-B_{jk}(t))=1\}\wedge\notag\\
        % &\hskip.5cm\mathds{1}\left\{\abs{y_{jk}(s_{jk}) - y_{jk}(t-B_{jk}(t))} \geq \epsilon\right\}
    \end{align}
    while under an OUTformation architecture, we use:
    \begin{align}\label{eq:out_\trigger_rule}
        &\gamma_{jk}^{(O)}(t,B_{jk}(t)) = \mathds{1}\{\beta_{jk}^{(O)}(t)=1\}\wedge \notag\\
        &\hskip.5cm\begin{cases}
            \mathds{1}\{ \abs{\hat{x}^{(b)}_{i_k}(t+B_{jk}(t)) - y_{jk}(t)} \geq \epsilon \} &\text{ if } s^{**} \geq s_{jk}\\
            \mathds{1}\left\{\abs{y_{jk}(s_{jk}) - y_{jk}(t)} \geq \epsilon\right\} &\text{ else}
        \end{cases}
    \end{align}
    Here, $s_{jk}{\,<\,}t$ is the time of sensor $j$'s previous transmission of component $k$, and $s^{**}{\,<\,}t{\,+\,}B_{jk}(t)$ is the time the broadcast value $\hat{x}^{(b)}_{i_k}(t{\,+\,}B_{jk}(t))$ was created.
    Threshold parameter $\epsilon > 0$ is chosen by design.
\end{definition}

\begin{notation}\label{nota:three_architectures}
    Under~\assum{paper_assumptions}, the three types of architectures we compare are specified below:
    \begin{enumerate}
        \item \textit{Pure INformation} architecture $IN_0$: each sensor $j$ implements \sensor{} rule~\eqn{in_\sensor_rule} with \sensor{} period $T_j^{(I)}$, and \trigger{} rule~\eqn{in_\trigger_rule} with zero threshold $\epsilon{\,=\,}0$.
    
        \item \textit{Absolute-Difference INformation} architecture $IN(\epsilon)$: each sensor $j$ implements \sensor{} rule~\eqn{in_\sensor_rule} with \sensor{} period $T_j^{(I)}$, and \trigger{} rule~\eqn{in_\trigger_rule} with fixed threshold $\epsilon{\,>\,}0$.
    
        \item \textit{OUTformation} architecture $OUT(\epsilon)$: each sensor $j$ implements \sensor{} rule~\eqn{out_\sensor_rule} with \sensor{} period $T_j^{(O)}$, \trigger{} rule~\eqn{out_\trigger_rule} with fixed threshold $\epsilon{\,>\,}0$, and the central processor uses the broadcast rule from~\defin{broadcast}.
    \end{enumerate}
\end{notation}

% The performance of each architecture is evaluated according to the following two metrics.

\begin{definition}[Rates per Component]\label{def:component_rates}
    For sensor $j$ under architecture $\chi{\,\in\,}\{I,O\}$, let $U_{jk}^{(\chi)}(s{\,:\,}t), D_{jk}^{(\chi)}(s{\,:\,}t){\,\in\,}\Nbb$ be the cumulative \textit{number of transmissions} and \textit{number of broadcasts received}, respectively, for component $k\in\{1,\cdots, m_j\}$ over the interval of time $[s,t)$.
    The \textit{communication delay} incurred per component $k$ is denoted $\Delta t^{(u)}\in\Rbb^{+}$ for uplink transmission, $\Delta t^{(d)}\in\Rbb^{+}$ for downlink broadcast.
    For each sensor, the \textit{rate of power} expended per component $k$ is $P_U\in\Rbb^{+}$ for transmission and $P_D\in\Rbb^{+}$ for receipt.
\end{definition}

\begin{definition}[Performance Metrics]\label{def:metrics}
    The \textit{mean-squared error (MSE)} of the environment state vector over experiment duration $[0,T_{\text{sim}})$ under architecture $\chi\in\{I,O\}$ is given by
    \begin{align}\label{eq:mse_metric}
        \frac{1}{T_{\text{sim}}}\sum\limits_{t=0}^{T_{\text{sim}}} \sum\limits_{i=1}^n\abs*{x_i(t) - \hat{x}^{(\chi)}_i(t)}^2
    \end{align}
    The \textit{total amount of power expenditure per sensor $j$} under architecture $\chi$ is given by:
    \begin{align}\label{eq:accum_power}
        &R_j^{(\chi)}(0:T_{\text{sim}}) = P_UU_j^{(\chi)}(0:T_{\text{sim}}) + P_DD_j^{(\chi)}(0:T_{\text{sim}})
    \end{align}
    where the power expenditure rates $P_U$,$P_D$ are from~\defin{component_rates}, and $U_j^{(\chi)}(s{\,:\,}t){\,\triangleq\,} \sum_{k=1}^{m_j}U_{jk}^{(\chi)}(s{\,:\,}t)$, likewise $D_j^{(\chi)}(s{\,:\,}t)\triangleq \sum_{k=1}^{m_j}D_{jk}^{(\chi)}(s{\,:\,}t)$.
\end{definition}

%%%%%%%%%%%%%%%%%%%%%%%%%%%%%%%%%%%%%%%%%%%%%%%%%%%%%%%%%%%%%%
% THEORETICAL ANALYSIS
%%%%%%%%%%%%%%%%%%%%%%%%%%%%%%%%%%%%%%%%%%%%%%%%%%%%%%%%%%%%%%

\section{Theoretical Analysis}\label{sec:analysis}
With the setup described above, we now characterize the tradeoff space between the MSE and sensor power expenditure metrics (see~\defin{metrics}) for the three architectures in~\nota{three_architectures}.
In this section, we introduce theory to analyze $IN(\epsilon)$ and $OUT(\epsilon)$ under the following simplified version
of the original problem setup in Section~\ref{subsec:problem_setup}.
Later, in~\sec{simulation}, we numerically study all three architectures for the original setup of Section~\ref{subsec:problem_setup}.
% Whether we are operating under the simplified version or the original setup, we show that similar characteristics arise in the performance metric tradeoff spaces for all three architectures.

\begin{setup}[Two Sensors over One-Component Change Environment]\label{set:periodic_bernoulli}
    We consider the true environment dynamics~\eqn{random_walk} when the vector of indicators $[\delta_1(c),\cdots,\delta_n(c)]^T$ can only take values from the set $\{\textbf{0}_n,\evect_1,\cdots,\evect_n,-\evect_1,\cdots,-\evect_n\}$ for all $c{\,\in\,}\Nbb$, where $\textbf{0}_n$ denotes the zero vector in $\Rbb^n$.
    Both types of architectures operate with $M{\,=\,}2$ sensors, where each sensor $j{\,\in\,}\{1,2\}$ has $m_j{\,\in\,}\Nbb$ components.
    Here, $m'{\,<\,}\min(m_1,m_2)$ components are shared in the sense of~\defin{shared} ($\Scal{\,\triangleq\,}\Scal_1{\,=\,}\Scal_2, \abs{\Scal}{\,=\,}m'$) and the remaining $m_j{\,-\,}m'$ are unshared ($\abs{\Ucal_j}{\,=\,}m_j{\,-\,}m'$).
    The sensors have the same sampling rate $\tau{\,\triangleq\,}t_1{\,=\,}t_2$ (see~\eqn{sensor}) such that $\Delta t/\tau{\,\in\,}\Nbb$, and use event-triggered \sensor{} $T_j^{(I)}{\,=\,}T_j^{(O)}{\,=\,}\tau$ (see~\defin{\sensor}).
    Because there are only two sensors in the network, sensor $1$ transmits immediately ($B_{1k}(t){\,=\,}0$) while sensor $2$ performs random backoff according to the cdf $F_B$ prescribed in~\assum{paper_assumptions}.
    % \blue{
    % Because the state $\xvect(t)$ undergoes at most one component change every $\Delta t$ timesteps, both sensors transmit all unshared components immediately, i.e. $B_{jk}(t){\,=\,}0$ for all $t$ and $k{\,\in\,}\Ucal_j$; for shared components $k{\,\in\,}\Scal$, sensor $1$ transmits immediately }
\end{setup}

\begin{notation}\label{nota:theory_notations}
    For each sensor $j\in\{1,2\}$ and each component $k\in\{1, \cdots, m_j\}$, let $p_{jk}^{(\chi)}(t,B_{jk}(t)) \triangleq \Pbb(\gamma_{jk}^{(\chi)}(t,B_{jk}(t)) = 1)$ be the probability that component $k$ is transmitted uplink by sensor $j$ at time $t$ under architecture $\chi$, using~\eqn{in_\trigger_rule} for $\chi{\,=\,}I$ and~\eqn{out_\trigger_rule} for $\chi{\,=\,}O$.
    Furthermore, for simplicity of expression, every interval $[c\Delta t, (c+1)\Delta t)$, where $c\in\Nbb$ and $\Delta t$ is from~\eqn{random_walk}, is referred to as \textit{environment interval $c$}.
\end{notation}
\begin{lemma}[Power from Unshared Components]\label{lem:power_unshared}
    For sensor $j{\,\in\,}\{1,2\}$ under~\setu{periodic_bernoulli}, the expected total power expended by unshared components over environment interval $c\in\Nbb$ is equivalent under $IN(\epsilon)$ or $OUT(\epsilon)$, and given by
    \begin{align}\label{eq:unshared_total_power}
        &\Ebb[R_{j,\Ucal_j}^{(\chi)}(c\Delta t:(c+1)\Delta t)] = P_U\sum\limits_{k\in\Ucal_j}\bigg(\sum\limits_{h=0}^{\Delta t/\tau}\notag\\
        &\hskip.5cm
        (1 - F_B(\Delta t - h\tau-\Delta t^{(u)}))p_{jk}^{(I)}(t_{c-1}(h),B_{jk}(t_{c-1}(h)))\notag\\
        &\hskip.5cm +
        \sum\limits_{h=1}^{\Delta t/\tau-1} F_B(\Delta t - h\tau-\Delta t^{(u)})p_{jk}^{(I)}(t_c(h),B_{jk}(t_c(h)))
        \bigg)
    \end{align}
    where $t_c(h)\triangleq c\Delta t+ht$ for any $c,h{\,\in\,}\Nbb$ is a placeholder, and $R_{j,\Ucal_j}^{(\chi)}(c\Delta t{\,:\,}(c+1)\Delta t)$ is the part of the power~\eqn{accum_power} contributed by unshared components.
\end{lemma}
\begin{proof}
    Under~\defin{broadcast}, no broadcasts are made for unshared components and~\eqn{out_\trigger_rule} is equivalent to~\eqn{in_\trigger_rule} when $k{\,\in\,}\Ucal_j$.
    Hence, $IN(\epsilon)$ and $OUT(\epsilon)$ have the same uplink communications for unshared components, and their contributed total power expenditures are equivalent.
    For placeholder $a{\,\in\,}\{c-1,c\}$, let $X_{jk}^{(\chi)}(a,h)$, $h{\,\in\,}\{0,\cdots, \Delta t/\tau-1\}$, be the indicator denoting whether or not $y_{jk}(a\Delta t + h\tau)$ is received during environment interval $c$ under architecture $\chi{\,\in\,}\{I,O\}$.
    By~\assum{paper_assumptions}, $y_{jk}((c{\,-\,}1)\Delta t {\,+\,}h\tau)$ is received during the interval with probability $1{\,-\,}F_B(\Delta t{\,-\,}h\tau{\,-\,}\Delta t^{(u)})$, and $y_{jk}(c\Delta t {\,+\,}h\tau)$ with probability $F_B(\Delta t{\,-\,}h\tau{\,-\,}\Delta t^{(u)})$.
    The result follows by definition of $p_{jk}^{(\chi)}(t,B_{jk}(t))$ in~\nota{theory_notations}
    % , from which we have $\Ebb[X_{jk}^{(\chi)}(c,h)]{\,=\,}p_{jk}^{(\chi)}(c\Delta t + h\tau, 0)$ 
    and $\Ebb[U_{jk}^{(\chi)}(c\Delta t{\,:\,}(c+1)\Delta t)] = \sum_{h=0}^{\Delta t/\tau} \Ebb[X_{jk}(c-1,h)]{\, +\,}\sum_{h=1}^{\Delta t/\tau-1} \Ebb[X_{jk}(c,h)]$.
\end{proof}

Because both $IN(\epsilon)$ and $OUT(\epsilon)$ have the same uplink communications for unshared components, the central processor has the same amount of data about the environment at each time $t$.
Thus, we also have the following result.

\begin{lemma}[MSE from Unshared Components]\label{lem:mse_unshared}
    Using the setup of~\lem{power_unshared}, suppose $\hat{\xvect}^{(\chi)}(t)$ is constructed via the fusion rule from~\assum{paper_assumptions} under $IN(\epsilon)$ when $\chi{\,=\,}I$ and $OUT(\epsilon)$ when $\chi{\,=\,}O$.
    Then
    \begin{align}
        \sum\limits_{i\in\Ncal}\abs*{x_i(t) - \hat{x}^{(I)}_i(t)}^2 = \sum\limits_{i\in\Ncal}\abs*{x_i(t) - \hat{x}^{(O)}_i(t)}^2
    \end{align}
    where $\Ncal{\,\triangleq\,}\{i_k{\,\in\,}\{1,\cdots,n\}\,|\,k{\,\in\,}\Ucal_1{\,\cup\,}\Ucal_2\}$ with $i_k$ denoting the full state vector component corresponding to component $k$ (see~\nota{full_vector_component}).
\end{lemma}
The more interesting comparison arises for shared components $k{\,\in\,}\Scal$.
% \revised{
To obtain concrete mathematical expressions, we present the next \underline{three} results over a specific chosen interval of communications $[T_0,T_f){\,\subset\,}[0,T_{\text{sim}})$ such that 1) $IN(\epsilon)$ and $OUT(\epsilon)$ begin from the same performance metric values at time $T_0$, and 2) the sample path of communications for shared component $k$ behaves exactly like the noiseless case.
% }
By deriving conditions of performance improvement for $OUT(\epsilon)$ over $IN(\epsilon)$ during $[T_0,T_f)$, we remark the implication that this improvement cascades over the longer interval $[T_0,T_{\text{sim}})$.

\begin{setup}[Interval of Communications I]\label{set:correct}
    Under~\setu{periodic_bernoulli}, choose the specific interval $[T_0,T_f)$ to be environment interval $c$ (i.e., $T_0{\,\triangleq\,}c\Delta t$, $T_f{\,\triangleq\,}(c+1)\Delta t$) such that the following occurs.
    The environment evolves such that $[\delta_1(c),\cdots,\delta_n(c)]=\evect_{i_k}$ for full state vector component $i_k{\,\in\,}\{1,\cdots,n\}$ defined in~\nota{full_vector_component} for a component $k{\,\in\,}\Scal$ which is shared in the sense of~\defin{shared}.
    The sample path of observations $\Ycal_k\triangleq\{y_{jk}(t), t\in[0,(c{\,+\,}1)\Delta t),j\in\{1,2\}\}$ for $k$ is such that for $\chi\in\{I,O\}$, $\beta_{jk}^{(\chi)}(c\Delta t){\,=\,}1$ and $\beta_{jk}^{(\chi)}(c\Delta t{\,+\,}h\tau){\,=\,}0$ for all $h{\,\leq\,}H-1$; here, $H{\,\in\,}\Nbb$ is defined in~\setu{periodic_bernoulli}, and $\beta_{jk}^{(\chi)}$ is the \sensor{} rule from~\defin{\sensor}.
\end{setup}

\begin{theorem}[Power from Shared Components]\label{thm:power_shared}
    % Under~\setu{periodic_bernoulli} and~\nota{theory_notations}, suppose $[\delta_1(c-1),\cdots,\delta_n(c-1)]=\evect_{i_k}$ for component $i_k{\,\in\,}\{1,\cdots,n\}$, 
    Suppose~\setu{correct} holds for shared component $k{\,\in\,}\Scal$ over environment interval $c{\,\in\,}\Nbb$ defined in~\nota{theory_notations}.
    % Further suppose the sensor observations $\Ycal{\,\triangleq\,}\{\yvect_j(t), t{\,\in\,}[c\Delta t,(c+1)\Delta t),j{\,\in\,}\{1,2\}\}$ yield communications during environment interval $c{\,\in\,}\Nbb$ which are correct in the sense of~\assum{correct}.
    %
    Then the expected difference between the power expended under $IN(\epsilon)$ and $OUT(\epsilon)$ is given by
    \begin{align}
        &\Ebb[R^{(I)}(c\Delta t:(c+1)\Delta t) - R^{(O)}(c\Delta t:(c+1)\Delta t)]\\
        &=
        (P_U - P_D)(1 - F_B(\Delta t^{(u)}+\Delta t^{(d)}))\Pbb(\abs{W_1-W_2}\geq \epsilon)\notag
    \end{align}
    where 
    % $\underline{d}$ is defined in~\eqn{random_walk}, 
    $W_1,W_2{\,\sim\,}\Ncal(0,\sigma^2)$ are independent random variables with $\sigma$ defined in~\eqn{sensor}, $\Delta t^{(u)}$, $\Delta t^{(d)}$, $P_U$, $P_D$ are the communication delays and power expenditure rates in~\defin{component_rates}, $F_B$ is the random backoff time distribution from~\setu{periodic_bernoulli}, and $R^{(\chi)}(c\Delta t{\,:\,}(c+1)\Delta t)$ is the power expended over both sensors.
\end{theorem}
\begin{proof}
    By the random backoff strategy of~\setu{periodic_bernoulli} and communications of~\setu{correct}, sensor $1$ transmits $y_{1k}(c\Delta t)$ at time $c\Delta t$ while sensor $2$ observes $y_{2k}(c\Delta t)$ and schedules to transmit at time $c\Delta t + B_{2k}(c\Delta t)$.
    Sensor $2$'s scheduled transmission is cancelled if only the central processor broadcasts $y_{1k}(c\Delta t)$ before time $c\Delta t + B_{2k}(c\Delta t)$, which occurs with probability $1-F_B(\Delta t^{(u)}+\Delta t^{(d)})$ under~\setu{periodic_bernoulli}, and if $\abs{y_{1k}(c\Delta t) - y_{2k}(c\Delta t)}{\,\geq\,} \epsilon$, which occurs with probability $\Pbb(\abs{W_1-W_2}{\,\geq\,}\epsilon)$ by~\eqn{sensor}.
    The result follows from the fact that $P_U-P_D$ is the difference between transmitting and broadcasting an extra component (see~\defin{component_rates}).
    % , and enumerating all possible $k\in\{1,\cdots, m'\}$.
\end{proof}

\begin{theorem}[MSE from Shared Components]\label{thm:mse_shared}
    Suppose~\setu{correct} holds for shared component $k{\,\in\,}\Scal$ over environment interval $c{\,\in\,}\Nbb$.
    Let $\hat{\xvect}^{(\chi)}(t)$ be constructed via the fusion rule from~\assum{paper_assumptions}, and suppose that $\hat{x}_{i_k}(c\Delta t) {\,\triangleq\,} \hat{x}_{i_k}^{(I)}(c\Delta t){\,=\,}\hat{x}_{i_k}^{(O)}(c\Delta t)$.
    Then the probability that the MSE contributed by component $i_k$, defined in~\nota{full_vector_component}, under $OUT(\epsilon)$ is less than that under $IN(\epsilon)$ is given by:
    \begin{align}\label{eq:mse_shared_prob}
        &(1 - F_B(\Delta t^{(u)}+\Delta t^{(d)}))\\
        &\hskip.5cm *\Pbb\left(\frac{1}{2}W_1W_2 + W_2^2 - \frac{3}{4}W_1^2 > 0,\hskip.1cm \abs{W_1-W_2}<\epsilon\right)\notag
    \end{align}
    where $\Delta t^{(u)}, \Delta t^{(d)}$ are the communication delays in~\defin{component_rates}, $F_B$ is the random backoff time distribution from~\setu{periodic_bernoulli}, 
    % $\epsilon>0$ is the threshold parameter from~\eqn{out_\trigger_rule}, 
    and $W_1,W_2{\,\sim\,}\Ncal(0,\sigma^2)$ are independent random variables with $\sigma$ from~\eqn{sensor}.
\end{theorem}
\begin{proof}
    For simplicity, we exclude the factor of $1/T_{\text{sim}}$ in~\eqn{mse_metric} throughout our proof.
    Under~\setu{correct}, the MSE of component $i_k$ during environment interval $c$ under $IN(\epsilon)$ is:
    \begin{align}\label{eq:mse_shared_in}
        &(\hat{x}_{i_k}(c\Delta t) - x_{i_k}(c\Delta t))^2\Delta t^{(u)}\notag\\
        &\hskip.7cm+ (y_{1k}(c\Delta t) - x_{i_k}(c\Delta t))^2(c\Delta t + B_{2k}(c\Delta t))\notag\\
        &\hskip.7cm+ \left(\frac{1}{2}(y_{1k}(c\Delta t) + y_{2k}(c\Delta t)) - x_{i_k}(c\Delta t)\right)^2\notag\\
        &\hskip2cm *(\Delta t - B_{2k}(c\Delta t) - \Delta t^{(u)})
    \end{align}
    Under $OUT(\epsilon)$, the MSE is the same as~\eqn{mse_shared_in} if sensor $2$ transmits $y_{2k}(t)$; this occurs when
    \begin{align}\label{eq:prob_sensor2_not_transmit_backoff}
        &\{B_{2k}(c\Delta t) < \Delta t^{(u)} + \Delta t^{(d)}\}\cup\\
        &\{B_{2k}(c\Delta t) \geq \Delta t^{(u)} + \Delta t^{(d)}, \abs{y_{1k}(c\Delta t) - y_{2k}(c\Delta t)}\geq\epsilon\}\notag
    \end{align}
    When sensor $2$ does not transmit, the MSE is:
    \begin{align}\label{eq:mse_shared_out}
        &(\hat{x}_{i_k}(c\Delta t) - x_{i_k}(c\Delta t))^2\Delta t^{(u)}\notag\\
        &\hskip.5cm+ (y_{1k}(c\Delta t) - x_{i_k}(c\Delta t))^2
        ((c+1)\Delta t - \Delta t^{(u)})
    \end{align}
    From~\eqn{sensor}, $(1/2)(y_{1k}(t)+y_{2k}(t)) - x_{i_k}(t) = (1/2)(W_1+W_2)$ for any $t$ and two independent $W_1,W_2\sim\Ncal(0,\sigma^2)$; likewise, $y_{1k}(t)- x_{i_k}(t) = W_1$.
    Hence, the difference between~\eqn{mse_shared_out} and~\eqn{mse_shared_in} yields $(\Delta t - B_{2k}(c\Delta t) - \Delta t^{(u)})((1/4)(W_1+W_2)^2 - W_1^2)$, which is positive under~\setu{periodic_bernoulli} if $(1/4)(W_1+W_2)^2{\,-\,}W_1^2{\,>\,}0$.
    Combining this with~\eqn{prob_sensor2_not_transmit_backoff} yields our result.
\end{proof}

\begin{figure}
    \begin{center}
        \includegraphics[width=0.98\columnwidth]{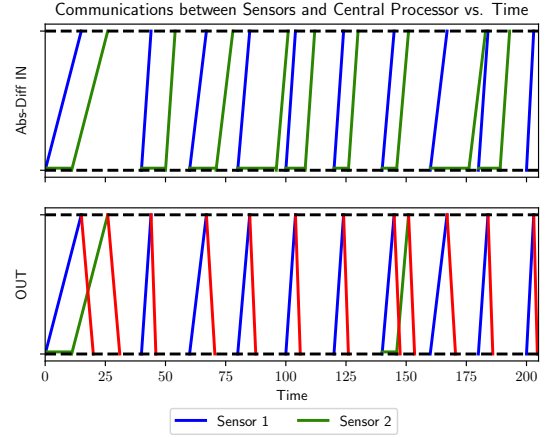}
    \end{center}
    \caption{A simple empirical illustration of the results of Theorems~\ref{thm:power_shared} and~\ref{thm:mse_shared} when $\sigma{\,=\,}0$ over environment~\eqn{random_walk}: time-evolution of the communications between a central processor and $M{\,=\,}2$ sensors under Absolute-Difference INformation architecture $IN(\epsilon)$ [Top] and OUTformation architecture $OUT(\epsilon)$ [Bottom] and event-triggered verification.
    Sensor $1$ transmissions are in blue, sensor $2$ in green, and central processor broadcasts in red.
    % Because~\eqn{in_\trigger_rule} yields the same result regardless of backoff time duration, and 
    Under $OUT(\epsilon)$, if the central processor broadcasts an update from sensor $1$, sensor $2$ may cancel its scheduled backoff transmission (e.g., $t{\,=\,}40,60,\cdots,120,160,\cdots$).
    The slopes of each line vary because communication delays are proportional to the number of components transmitted/broadcast (see~\defin{component_rates}).
    }
    \label{fig:event_triggered}
\end{figure}

\begin{remark}[Implications of Theorems~\ref{thm:power_shared} and~\ref{thm:mse_shared}]\label{rmk:event_triggered}
    In~\fig{event_triggered}, we demonstrate the cascading effect of the performance improvements derived in Theorems~\ref{thm:power_shared} and~\ref{thm:mse_shared} by considering a longer interval of time than~\setu{correct}.
    For significance, the original setting of multiple simultaneous component changes (see Section~\ref{subsec:problem_setup}) with no measurement noise $(\sigma{\,=\,}0)$ is plotted.
    For a general number $M{\,\geq\,}2$ of sensors with small $\sigma$ in~\eqn{sensor}, multiple sensors may decrease in both uplink and downlink power at the expense of additional downlink power for one sensor, and~\thm{power_shared} suggests an overall decrease in total power expenditure using $OUT(\epsilon)$ over $IN(\epsilon)$.
    Consequently,~\thm{mse_shared} and~\fig{event_triggered} suggest that the difference in MSE performance between $OUT(\epsilon)$ and $IN(\epsilon)$ is always zero when $\sigma = 0$ in~\eqn{sensor}; when $\sigma\neq 0$, $OUT(\epsilon)$ yields less MSE than $IN(\epsilon)$ with probability~\eqn{mse_shared_prob}.
\end{remark}
An interesting MSE advantage of $OUT(\epsilon)$ over $IN(\epsilon)$ arises when we use larger verification periods $T_j^{(I)}{\,>\,}\tau_j$, defined in~\defin{\sensor}.
% \revised{
Here, instead of~\setu{correct}, we use the following interval of communications instead.
% $[T_0,T_f){\,\subset\,}[0,T_{\text{sim}})$, then remark the implications of cascading improvement over the longer interval $[T_0,T_{\text{sim}})$.
% }

\begin{setup}[Interval of Communications II]\label{set:single_interval}
    % We consider the original dynamics~\eqn{random_walk} with nonzero random backoff time for both sensors (see~\assum{paper_assumptions}).
    % For sampling time $\tau_j$ defined in~\eqn{sensor} and \sensor{} periods defined in~\defin{\sensor}, let $T_j^{(I)}{\,=\,}T_j^{(O)}{\,=\,}T_j{\,>\,}\tau_j$ for sensor $j{\,\in\,}\{1,2\}$.
    % Choose interval $\Ical\triangleq[T_0,T_f)$, $T_0{\,\triangleq\,}(a_2{\,-\,}1)T_2$ and $T_f{\,\triangleq\,}(a_1{\,+\,}1)T_1$ for some $a_1,a_2{\,\in\,}\Nbb$ such that $(a_2{\,-\,}1)T_2 < a_1T_1 < a_2T_2 < (a_1{\,+\,}1)T_1$.
    % The environment~\eqn{random_walk} evolves such that for each sensor $j\in\{1,2\}$, 1) one unshared component $k_j\in\Ucal_j$ changes value once by magnitude $d_j$ during $((a_j-1)T_j, a_jT_j]$, and 2) one shared component $k'\in\Scal$ changes value once by magnitude $d'$ during $((a_2-1)T_2, a_1T_1]$ and remains constant during $(a_1T_1, a_2T_2]$; here, $d_1,d_2,d'{\,\sim\,}\Ucal[\underline{d},\overline{d}]$ via the stepsize distribution from~\eqn{random_walk} and, using~\nota{full_vector_component}, no other components $i\in\{1,\cdots, n\}/\{i_{k_1}, i_{k_2}, i_{k'}\}$ change value over $\Ical$.
    % The sample path of observations $\Ycal\triangleq\{\yvect_j(t), t{\,\in\,}[0,(a_2-1)T_2)\cup\Ical,j{\,\in\,}\{1,2\}\}$ is such that $\beta_{2k_2}^{(\chi)}((a_2-1)T_2){\,=\,}1$, $\beta_{1k_1}^{(\chi)}(a_1T_1){\,=\,}1$, and $\beta_{jk'}^{(\chi)}(a_jT_j){\,=\,}1$ for $j\in\{1,2\}$, where $\beta_{jk}^{(\chi)}$ is the \sensor{} rule from~\defin{\sensor}.
    %
    % \revised{
    We consider the original dynamics~\eqn{random_walk} with 
    % possibly multiple simultaneous component changes 
    nonzero random backoff time for both sensors (see~\assum{paper_assumptions}).
    For sampling time $\tau_j$ defined in~\eqn{sensor} and \sensor{} periods defined in~\defin{\sensor}, let $T_j^{(I)}{\,=\,}T_j^{(O)}{\,=\,}T_j{\,>\,}\tau_j$ for sensor $j{\,\in\,}\{1,2\}$.
    Choose interval $[T_0,T_f)$, $T_0{\,\triangleq\,}(a_1{\,-\,}1)T_1$ and $T_f{\,\triangleq\,}a_1T_1$ such that $(a_1-1)T_1{\,<\,}a_2T_2{\,<\,}a_1T_1$ for some $a_1,a_2{\,\in\,}\Nbb$.
    The environment~\eqn{random_walk} evolves such that 1) one shared component $k'\in\Scal$ changes value once by magnitude $d'$ during $((a_1-1)T_1, a_2T_2]$ and remains constant during $(a_2T_2, a_1T_1]$, and 2) one unshared component $k_1\in\Ucal_1$ of sensor $1$ remains constant during $((a_1-1)T_1, a_2T_2]$ and changes value once by magnitude $d_1$ during $(a_2T_2, a_1T_1]$, and
    3) no other full-state components $i\in\{1,\cdots, n\}/\{i_{k_1}, i_{k'}\}$ (see~\nota{full_vector_component}) change value over $[T_0,T_f)$.
    Here, $d_1,d'{\,\sim\,}\Ucal[\underline{d},\overline{d}]$ via the stepsize distribution from~\eqn{random_walk}.
    The sample path of observations $\Ycal\triangleq\{\yvect_j(t), t{\,\in\,}[0,(a_1-1)T_1)\cup[T_0,T_f),j{\,\in\,}\{1,2\}\}$ is such that $\beta_{1k_1}^{(\chi)}(a_1T_1){\,=\,}1$ and $\beta_{jk'}^{(\chi)}(a_jT_j){\,=\,}1$ for $j\in\{1,2\}$, where $\beta_{jk}^{(\chi)}$ is the \sensor{} rule from~\defin{\sensor}.
    % }
\end{setup}

\begin{theorem}[MSE with Longer Verification Period]\label{thm:mse_shared_gen}
    Consider the interval defined in~\setu{single_interval}.
    Let $\xvect^{(\chi)}(t)$ be constructed via the fusion rule from~\assum{paper_assumptions} such that $\hat{x}_{i_k}(T_0) \triangleq \hat{x}_{i_k}^{(I)}(T_0) = \hat{x}_{i_k}^{(O)}(T_0)$ where $T_0$ and $k\in\{k_1,k'\}$ are defined in~\setu{single_interval}.
    Then the MSE contributed by component $i_{k_1}$ under $OUT(\epsilon)$ is less than that under $IN(\epsilon)$ during interval $[T_0,T_f)$ with probability:
    \begin{align}\label{eq:mse_shared_gen}
        &\Pbb\left(\abs{d_1}>0, \abs{W_2' - W_1'} < \epsilon\right)*\\
        &\hskip.5cm(1 - G_{B_1,B_2}(a_1T_1 - a_2T_2 + \Delta t^{(u)} + \Delta t^{(d)})\notag
        % &\Pbb\left(W_2^2 - (d_2+W_1)^2 \leq 0, -\epsilon-d' < W_2' - W_1' < \epsilon - d'\right)\notag
    \end{align}
    Here, $G_{B_1,B_2}(s){\,\triangleq\,}\Pbb(B_1{\,-\,}B_2 {\,\leq\,}s)$ for two independent $B_1,B_2$ distributed according to $F_B$ from~\assum{paper_assumptions}, 
    % $d_1,d_2,d'$ are defined in~\setu{single_interval},
    $W_1,W_1',W_2,W_2'\sim\Ncal(0,\sigma^2)$ are independent random variables with $\sigma>0$ from~\eqn{sensor}, and the communication delays $\Delta t^{(u)}, \Delta t^{(d)}$ are in~\defin{component_rates}.
    % and $\epsilon>0$ is the threshold parameter from~\eqn{out_\trigger_rule}.
\end{theorem}
\begin{proof}
    As in the proof of~\thm{mse_shared}, we exclude the factor of $1/T_{\text{sim}}$ from~\eqn{mse_metric}.
    Under $OUT(\epsilon)$ in~\setu{single_interval}, sensor $1$ receives a broadcast update $y_{2k'}(a_2T_2)$ about component $k'$ if
    \begin{align}\label{eq:prob_sensor2_not_transmit_backoff_gen}
        &\{B_{2k'}(a_2T_2) - B_{1k'}(a_1T_1) \geq \notag\\
        &\hskip1cm a_1T_1 - a_2T_2 + \Delta t^{(u)} + \Delta t^{(d)}\}\cap\notag\\
        &\hskip2cm \{\abs{y_{2k'}(a_2T_2) - y_{1k'}(a_1T_1)} < \epsilon\}
    \end{align}
    If~\eqn{prob_sensor2_not_transmit_backoff_gen} holds, sensor $1$ only transmits for component $k_1$.
    Under $IN(\epsilon)$, sensor $1$ transmits both components $k'$ and $k_1$.
    The central processor receives $y_{1k_1}(a_1T_1)$ with an additional uplink delay of $\Delta t^{(u)}$ compared to $OUT(\epsilon)$.
    Thus, the difference in MSE of component $i_{k_1}$ between $OUT(\epsilon)$ and $IN(\epsilon)$ during interval $[T_0,T_f)$ is:
    \begin{align}\label{eq:mse_shared_gen_diff}
        &\Delta t^{(u)}((y_{1k_1}(a_1T_1) - x_{i_{k_1}}(c\Delta t))^2\notag\\
        &\hskip1cm- (\hat{x}_{i_{k_1}}(a_1T_1) - x_{i_{k_1}}(c\Delta t))^2)
    \end{align}
    % \begin{align}\label{eq:mse_shared_gen_in}
    %     &\sum\limits_{c=\floor{a_1T_1/\Delta t}}^{\floor{(a_2T_2 + B_2(a_2T_2) + \Delta t^{(u)})/\Delta t}}(\hat{x}_{i_{k_2}}(a_1T_1) - x_{i_{k_2}}(c\Delta t))^2\notag\\
    %     &*\left(\max{a_1T_1,c\Delta t}, \min(a_2T_2 + B_2(a_2T_2) + \Delta t^{(u)},(c+1)\Delta t)\right)\notag\\
    %     &+\sum\limits_{c=\floor{(a_2T_2 + B_2(a_2T_2) + \Delta t^{(u)})/\Delta t}}^{\floor{(a+1)T_1/\Delta t}}(y_{2k_2}(a_2T_2) - x_{i_{k_2}}(c\Delta t))^2
    % \end{align}
    where $c{\,\in\,}\Nbb$ is such that $c\Delta t{\,\leq\,}a_1T_1$ is the time of the last true change in component $k_1$ which was not transmitted to the central processor, with
    % such that $[a_2T_2+B_2(a_2T_2)+\Delta t^{(u)}, a_2T_2+B_2(a_2T_2)+2\Delta t^{(u)}] \subset [c\Delta t, (c+1)\Delta t)$,
    $\Delta t$ defined in~\eqn{random_walk}.
    By~\setu{single_interval} and~\eqn{sensor}, we have $y_{1k_1}(a_1T_1) - x_{i_{k_1}}(c\Delta t) = W_1$, $\hat{x}_{i_{k_1}}(a_1T_1) - x_{i_{k_1}}(c\Delta t) = d_1 + W_1$, and $y_{2k'}(a_2T_2) - y_{1k'}(a_1T_1) = W_2' - W_1'$ for some independent $W_1,W_1',W_2'{\,\sim\,}\Ncal(0,\sigma^2)$.
    Combined with~\eqn{prob_sensor2_not_transmit_backoff_gen}, the result follows.
\end{proof}

\begin{figure}
    \begin{center}
        \includegraphics[width=0.98\columnwidth]{../../figures/time_comms}
        \includegraphics[width=0.9\columnwidth]{../../figures/states}
    \end{center}
    \caption{[Top] A version of~\fig{event_triggered} under original dynamics~\eqn{random_walk} with longer \sensor{} periods $T_1^{(\chi)}{\,=\,}23, T_2^{(\chi)}{\,=\,}41$, $\chi\in\{I,O\}$.
    All three architectures $IN_0$, $IN(\epsilon)$, and $OUT(\epsilon)$ of~\nota{three_architectures} are plotted.
    Note three times where the uplink communications of $OUT(\epsilon)$ are less than $IN(\epsilon)$: 1) at $t{\,=\,}92, 138$, where sensor $1$ cancels its scheduled transmission
    % due to a broadcast created by sensor $2$'s transmission at $t{\,=\,}82$, 
    and 2) at time $t{\,=\,}161$, where the slope of sensor $1$'s uplink transmission is steeper.
    % due to a broadcast created by sensor $2$'s transmission at $t{\,=\,}123$.
    This suggests that updates from sensor observations are received more quickly on average under $OUT(\epsilon)$ than $IN(\epsilon)$.
    % sigma = 0
    % # T_s_list_IN = [23.0, 41.0]; shift_list_IN = [0.0, 0.0]. T_s_list_OUT = [23.0, 41.0]; shift_list_OUT = [0.0, 0.0]
    [Bottom] The evolution of components $7,10,12$ of $\xvect(t)$ and $\hat{\xvect}^{(\chi)}(t)$ over time.
    Since the central processor receives transmissions more quickly, 
    % $OUT(\epsilon)$ is more sensitive to changes in the environment than $IN(\epsilon)$ and 
    $\hat{\xvect}^{(O)}(t)$ tracks $\xvect(t)$ more accurately than $\hat{\xvect}^{(I)}(t)$.
    % sigma = 0.0
    % Delta_t = 20.0 
    % T_s_list_IN = [30.0, 30.0]; shift_list_IN = [0.0, 15.0] # list of sensor transmission periods and shift offsets for IN strats. T_s_list_OUT = [30.0, 30.0]; shift_list_OUT = [0.0, 15.0] # list of sensor transmission periods and shift offsets for OUT strats
    }
    \label{fig:time_triggered}
\end{figure}

\begin{figure*}
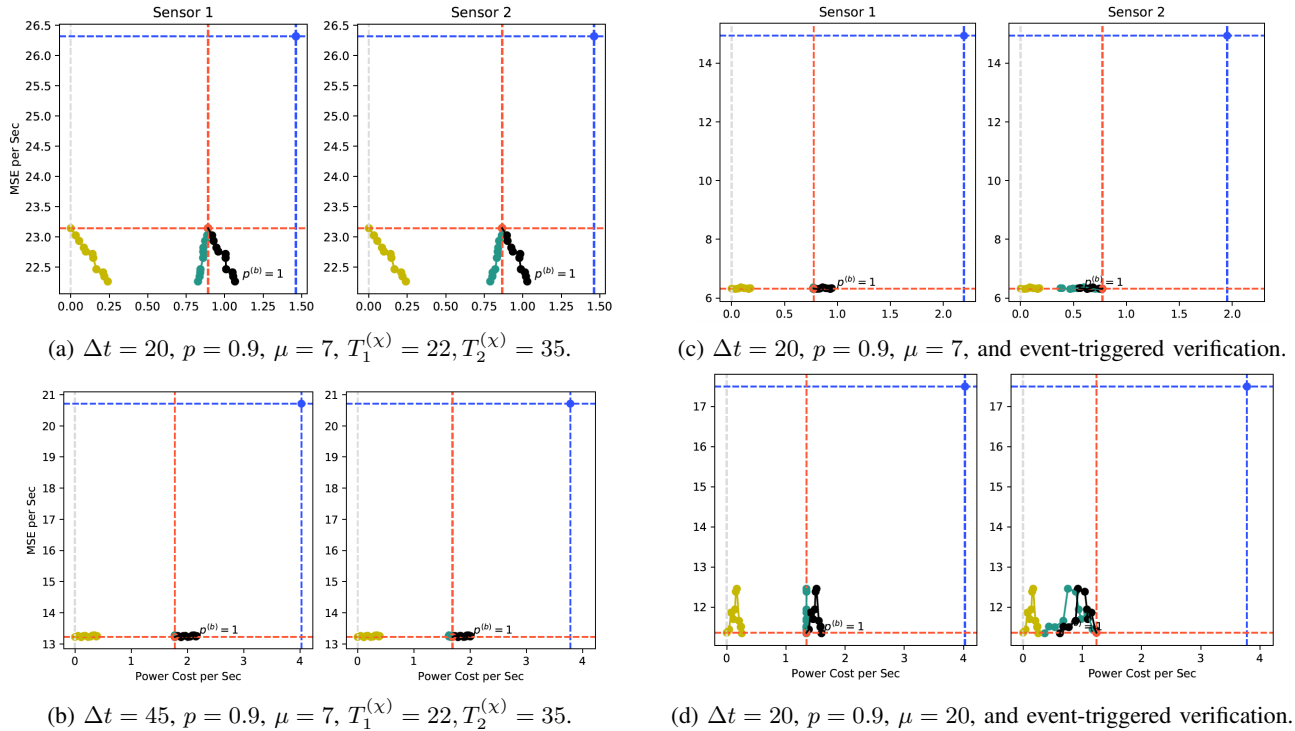

  \centering
  \begin{tabular}{ c @{\hspace{20pt}} c }
    \includegraphics[width=0.95\columnwidth]{../figures/final_figures/top.png} &
      \includegraphics[width=0.9\columnwidth]{../figures/final_figures/mid_top.png}\\
    \small (a) $\Delta t = 20$, $p=0.9$, $\mu=7$, $T_1^{(\chi)} = 22, T_2^{(\chi)} = 35$. & \small (c) $\Delta t = 20$, $p=0.9$, $\mu=7$, and event-triggered \sensor.\\
    \includegraphics[width=0.9\columnwidth]{../figures/final_figures/mid_bottom.png} &
      \includegraphics[width=0.92\columnwidth]{../figures/final_figures/bottom.png}\\
      \small (b) $\Delta t = 45$, $p=0.9$, $\mu=7$, $T_1^{(\chi)} = 22, T_2^{(\chi)} = 35$. & \small (d) $\Delta t = 20$, $p=0.9$, $\mu=20$, and event-triggered \sensor.
  \end{tabular}
  \medskip
    \caption{Plots of the MSE per timestep against power expenditure per timestep for both sensors $1$ and $2$, averaged over $40$ Monte-Carlo simulations of environment~\eqn{random_walk}.
    Each row of subfigures differ by their values of $\Delta t$ and $p$ (see~\eqn{random_walk}), mean random backoff time $\mu$, and frequency of verification (see~\defin{\sensor}).
    The architectures compared are described in~\nota{three_architectures}.
    The blue dot corresponds to $IN_0$, and the orange dot for $IN(\epsilon)$, with dashed lines to emphasize their visibility.
    For $OUT(\epsilon)$, a constant broadcast probability $p^{(b)}$ is ranged over $\{0,0.1, \cdots, 1\}$, and we make separate curves for the uplink, downlink, and the total (uplink $+$ downlink) power expenditures; gold dots for downlink, green for uplink, and black for the combined total.
    As in~\remk{time_triggered}, $IN_0$ performs the worst in both metrics for all four cases.
    The green and black dots for $OUT(\epsilon)$ with $p^{(b)}{\,=\,}0$ coincide with $IN(\epsilon)$ because no broadcasts are made.
    Under environments with fast variation (small $\Delta t$) relative to the \sensor{} periods $T_j^{(\chi)}$, $OUT(\epsilon)$ detects changes more quickly than $IN(\epsilon)$, providing a reduction in MSE at additional downlink power (see (a) and~\remk{time_triggered}); this tradeoff becomes dampened when $\Delta t$ is large, and the MSE performance rarely varies with power expenditure (see (b)).
    Under event-triggered verification, $OUT(\epsilon)$ maintains similar MSE as $IN(\epsilon)$ by having one sensor reduce power expenditure at the cost of the other sensor (see (c) and~\remk{event_triggered}); longer mean backoff time causes increasingly outdated data to be received, which increases the MSE (see (d)).
    % More detailed insights are provided in~\sec{simulation}.
    }
    \label{fig:all_tradeoff_curves}
\end{figure*}

\begin{remark}[Implications of~\thm{mse_shared_gen}]\label{rmk:time_triggered}
    In~\fig{time_triggered}, we demonstrate the cascading effect of the MSE advantage derived in~\thm{mse_shared_gen} by considering a longer interval of time than~\setu{single_interval}.
    For presentation clarity, we choose prime-valued $T_j^{(\chi)}$ and zero backoff time.
    For comparison, we include the communications and MSE for the pure INformation architecture $IN_0$ (see~\nota{three_architectures}); $IN_0$ transmits the most number of components uplink, consequently incurring the longest transmission delays and largest MSE among the three architectures.
    By \trigger{} rule~\eqn{out_\trigger_rule}, $OUT(\epsilon)$ always transmits the same number of components or less than $IN(\epsilon)$.
    Less components are likely to be transmitted for small measurement noise $\sigma$ relative to the distribution of the stepsize $\dvect(c), c\in\Nbb$ in~\eqn{random_walk}, since one sensor cancels backoff transmissions if broadcast updates from the other sensor contain similar component values (e.g., $t{\,=\,}92, 138,161$ in~\fig{time_triggered}).
    Because $OUT(\epsilon)$ transmits less uplink packets on average among the three architectures, it also incurs the least uplink delays, and so the central processor under $OUT(\epsilon)$ detects changes in the environment state most quickly.
    % This is also captured by~\thm{mse_shared_gen}, where the event $\{W_2^2 - (d_2+W_1)^2 \leq 0\}$ occurs with high probability.
    On the other hand, for large $\sigma$ or a suboptimal choice of $\epsilon$, the event $\abs{W_2' - W_1'} < \epsilon$ occurs with low probability.
    % Under event-triggered \sensor{}, the implications of~\thm{mse_shared} arise similarly.
\end{remark}

%%%%%%%%%%%%%%%%%%%%%%%%%%%%%%%%%%%%%%%%%%%%%%%%%%%%%%%%%%%%%%
% NUMERICAL SIMULATIONS
%%%%%%%%%%%%%%%%%%%%%%%%%%%%%%%%%%%%%%%%%%%%%%%%%%%%%%%%%%%%%%

\section{Numerical Analysis}\label{sec:simulation}
In this section, we supplement the theoretical insights derived from the previous~\sec{analysis} by empirically plotting the tradeoff space under a variety of dynamics and parameter choices.
In particular, we implement the original setup of Section~\ref{subsec:problem_setup}, consider longer intervals than those specified in Settings~\ref{set:correct} and~\ref{set:single_interval}, and remove specific constraints on the sample path of observations $\Ycal\triangleq\{\yvect_j(t), t{\,\in\,}[0,T_{\text{sim}}),j{\,\in\,}\{1,2\}\}$ and state estimates $\hat{\xvect}^{(\chi)},\chi{\,\in\,}\{I,O\}$ imposed in Theorems~\ref{thm:power_shared},~\ref{thm:mse_shared}, and~\ref{thm:mse_shared_gen}.
% white noise variance $\sigma = 0.1$, communication delays per component $\Delta t^{(u)} = 2.0, \Delta t^{(d)} = 0.5$, and power expenditure rates $P_U = 5.0, P_D = 1.0$.
We consider the effects of varying three parameters which demonstrate the most interesting comparisons between the performance of the three architectures from~\nota{three_architectures}: 1) the variation in the environment $\Delta t$ (see~\eqn{random_walk}), 2) the mean $\mu$ of the random backoff time distribution $F_B$ (see~\assum{paper_assumptions}), and 3) the frequency of verification (see~\defin{\sensor}).
For $OUT(\epsilon)$, we additionally vary over $p^{(b)}\in[0,1]$, a constant probability in which the central processor makes a broadcast upon receiving a transmission.
We choose communication delays $\Delta t^{(d)}{\,<\,}\Delta t^{(u)}$ and power expenditure rates $P_U{\,<\,}P_D$ (see~\defin{component_rates}).
% , and use a fortunate choice of threshold $\epsilon{\,<\,}\underline{d}$ for $\underline{d}$ (whose value is unknown to all three architectures) is defined in~\eqn{random_walk}.
% , and \trigger{} rule threshold values $\epsilon<\underline{d}$.
% These and other parameters (e.g. white noise variance $\sigma$) may be varied as well, but they are kept fixed for this experiment.

The results are demonstrated in~\fig{all_tradeoff_curves}.
With a larger number $M\geq 2$ of sensors, one may be inclined to believe that $IN_0$ obtains the lowest MSE under an averaging fusion rule (see~\assum{paper_assumptions}) by the law of large numbers, because $IN_0$ transmits the largest number of independent, redundant noisy observations.
However, when communication delays and random backoff times are involved, a larger number of transmissions will take a longer delay to be received by the central processor, causing an increase in MSE.
This effect was demonstrated in~\remk{time_triggered}.
% This phenomenon also occurs for $M\geq 2$: a larger number of sensors accumulating a larger number of independent observations does not automatically lead to a smaller MSE.
% The remainder of our discussion compares $IN(\epsilon)$ and $OUT(\epsilon)$.
% , the two most interesting architectures among the three in~\nota{three_architectures}.

%(a)
For the values chosen to generate~\fig{all_tradeoff_curves} (a), $OUT(\epsilon)$ is greater than $IN(\epsilon)$ in power expenditure while $IN(\epsilon)$ is greater than $OUT(\epsilon)$ in MSE for all $p^{(b)}\in(0,1]$.
This effect was demonstrated in~\remk{time_triggered};
% for small measurement noise $\sigma$ over environments which are fast relative to the \sensor{} periods (i.e., $\Delta t < T_j^{(\chi)}, \chi\in\{I,O\}$),
$OUT(\epsilon)$ tracks changes in the true environment more quickly than $IN(\epsilon)$, but with greater downlink power expenditure.
In~\fig{all_tradeoff_curves} (a), this tradeoff is shown explicitly: increasing downlink power (in gold) allows for decreasing uplink power (in green) for both sensors.
This suggests that for $OUT(\epsilon)$ applied to settings similar to~\fig{all_tradeoff_curves} (a), a balance between total power expenditure and MSE may be obtained by choosing $p^{(b)}$ appropriately.
%
% (b)
On the other hand, in~\fig{all_tradeoff_curves} (b), transmissions are more likely to be received by the central processor while the environment state has not changed (i.e., $T_j^{(\chi)} + \mu < \Delta t, \chi\in\{I,O\}$).
% In this setting, $IN(\epsilon)$ performs just as well in MSE as $OUT(\epsilon)$ .
% This is because the central processor is highly likely to receive updated transmissions from either sensor in between consecutive environment changes.
Essentially, if the environment variation is slow relative to the sensors' \sensor{} periods, then $OUT(\epsilon)$ maintains approximately the same MSE as $IN(\epsilon)$ on average, aside from small variations due to noise $\sigma$, and wastes power by broadcasting unnecessarily.

% (c)
The tradeoff space in~\fig{all_tradeoff_curves} (c) demonstrates that~\remk{event_triggered} holds more generally for slower-changing environments (large $\Delta t$) and small random backoff time relative to the environment (i.e., $\mu{\,<\,}\Delta t$).
% Compared to $IN(\epsilon)$,
For $OUT(\epsilon)$ under small $\sigma$, sensor $1$ expends power for both transmitting and receiving broadcasts, while sensor $2$ expends mostly downlink power for receiving broadcast updates.
% which often cancel its scheduled backoff transmissions.
Because both sensors operate on event-triggered verification, $OUT(\epsilon)$ is, on average, the same as $IN(\epsilon)$ in the MSE; any variations such that $OUT(\epsilon)$ obtains more or less MSE than $IN(\epsilon)$ is a result of noise. 
% Thus, under small $\sigma$,
% a backed-off transmission from sensor $2$ contains approximately the same data as an immediate transmission from sensor $1$,
% there is not much difference in MSE between $OUT(\epsilon)$ and $IN(\epsilon)$ by terminating sensor $2$'s transmission.
This suggests that for $OUT(\epsilon)$ applied to settings similar to~\fig{all_tradeoff_curves} (c), a balance for the total power expenditure may be obtained by choosing $p^{(b)}$ appropriately, at little change in MSE.
%
% (d)
When $\mu \geq \Delta t$, as in~\fig{all_tradeoff_curves} (d), the power expenditure follows the same trends as when $\mu<\Delta t$, but the overall MSE tends to be larger.
This is expected because $\mu{\,>\,}\Delta t$ implies that on average, sensor $2$'s observations about $\xvect(c\Delta t)$ are received when the true state is already changed to $\xvect((c+1)\Delta t)$.
% , causing the MSE to grow large.
% Because the nature of the broadcast rule (\defin{broadcast}) is to ignore outdated transmissions, $OUT(\epsilon)$ becomes more sensitive to measurement noise $\sigma$, potentially increasing the MSE.

%% OTHER COMPARISONS
%
% \item Overall, for event-triggered \sensor{}, an OUTformation architecture can consistently reach near the same level of MSE performance as an INformation architecture with less power expenditure, provided $P_U << P_D$ and $\sigma$ is small relative to the magnitude of changes in the true environment.
%
% \item $\Delta t^{(u)} < \Delta t^{(d)}$: This is expected since $OUT(\epsilon)$ behaves exactly like $IN(\epsilon)$ when the broadcast is more outdated than the previous transmission, which occurs more often with $\Delta t^{(u)} < \Delta t^{(d)}$.
% Hence, even though $OUT(\epsilon)$ spends more total power than $IN(\epsilon)$, the MSE of $OUT(\epsilon)$ never exceeds that of $IN(\epsilon)$.

%%%%%%%%%%%%%%%%%%%%%%%%%%%%%%%%%%%%%%%%%%%%%%%%%%%%%%%%%%%%%%
% CONCLUSION
%%%%%%%%%%%%%%%%%%%%%%%%%%%%%%%%%%%%%%%%%%%%%%%%%%%%%%%%%%%%%%

\section{Conclusion}
We provided a theoretical and numerical characterization of the tradeoff space for architectures with broadcast feedback (OUTformation) and architectures without feedback (INformation) using two performance metrics: the mean-squared error of the central processor's estimate of the environment state and the total power expenditure per sensor.
Our study was motivated towards enabling users to make informed design choices in distributed data-gathering architectures for large-scale network environments under constraints such as nonzero communication delays and limited knowledge of the environment dynamics.
We found that under an event-triggered verification rule (see~\defin{\sensor}), OUTformation architectures expend less uplink power on average than INformation, and yields similar MSE when variation in the environment is large relative to mean backoff time
% ; this potentially allows for less total power expenditure in multiple sensors at the expense of additional downlink power in a select few
(see~\thm{power_shared},~\remk{event_triggered},~\fig{all_tradeoff_curves} (c) and (d)).
% Depending on the relationship between $\mu$ and $\Delta t$, the MSE under OUTformation is able to match that of INformation (see).
%
Under a periodic verification rule, OUTformation architectures enable a reduced MSE on average compared to INformation, but with additional downlink power that potentially increases the total power expenditure (see~\thm{mse_shared},~\thm{mse_shared_gen}, and~\remk{time_triggered}).
This suggests that by varying parameters (i.e., the broadcast probability $p^{(b)}$ described in~\sec{simulation}), the OUTformation architecture can attain a better tradeoff between the MSE and the total power expenditure (see~\fig{all_tradeoff_curves} (a) and (b)).
In conclusion, the main advantage of feedback is that each sensor's understanding of the central processor's state estimate improves.
Theoretical and numerical analysis of performance tradeoffs for architectures with and without feedback when other parameters (e.g., threshold $\epsilon$, number of sensors) are varied is a natural subject of future work.
\bibliographystyle{IEEEtran}
\bibliography{%
bibi%
}

\begin{thebibliography}{1}
\providecommand{\url}[1]{#1}
\csname url@samestyle\endcsname
\providecommand{\newblock}{\relax}
\providecommand{\bibinfo}[2]{#2}
\providecommand{\BIBentrySTDinterwordspacing}{\spaceskip=0pt\relax}
\providecommand{\BIBentryALTinterwordstretchfactor}{4}
\providecommand{\BIBentryALTinterwordspacing}{\spaceskip=\fontdimen2\font plus
\BIBentryALTinterwordstretchfactor\fontdimen3\font minus
  \fontdimen4\font\relax}
\providecommand{\BIBforeignlanguage}[2]{{%
\expandafter\ifx\csname l@#1\endcsname\relax
\typeout{** WARNING: IEEEtran.bst: No hyphenation pattern has been}%
\typeout{** loaded for the language `#1'. Using the pattern for}%
\typeout{** the default language instead.}%
\else
\language=\csname l@#1\endcsname
\fi
#2}}
\providecommand{\BIBdecl}{\relax}
\BIBdecl

\bibitem{castanedo13}
F.~Castanedo, ``A review of data fusion techniques,'' \emph{The Scientific
  World Journal}, vol. 2013, pp. 1--19, Oct 2013.

\bibitem{liggins97}
M.~Liggins, C.-Y. Chong, I.~Kadar, M.~Alford, V.~Vannicola, and S.~Thomopoulos,
  ``Distributed fusion architectures and algorithms for target tracking,''
  \emph{Proceedings of the IEEE}, vol.~85, no.~1, pp. 95--107, 1997.

\bibitem{bandyopadhyay03}
S.~Bandyopadhyay and E.~J. Coyle, ``An energy efficient hierarchical clustering
  algorithm for wireless sensor networks,'' in \emph{IEEE INFOCOM 2003.},
  vol.~3, 2003, pp. 1713--1723.

\bibitem{younis04}
O.~{Younis} and S.~{Fahmy}, ``Distributed clustering in ad-hoc sensor networks:
  a hybrid, energy-efficient approach,'' in \emph{IEEE INFOCOM 2004.}, vol.~1,
  2004.

\bibitem{chen15}
B.~Chen, W.-A. Zhang, L.~Yu, G.~Hu, and H.~Song, ``Distributed fusion
  estimation with communication bandwidth constraints,'' \emph{IEEE
  Transactions on Automatic Control}, vol.~60, no.~5, pp. 1398--1403, 2015.

\bibitem{ge14outformation}
X.~Ge and Q.-L. Han, ``Distributed event-triggered {$H_{\infty}$} filtering
  over sensor networks with communication delays,'' \emph{Information
  Sciences}, vol. 291, pp. 128 -- 142, 2014.

\bibitem{nutt82}
G.~Nutt and D.~Bayer, ``{Performance of CSMA/CD Networks Under Combined Voice
  and Data Loads},'' \emph{IEEE Transactions on Communications}, vol.~30,
  no.~1, pp. 6--11, 1982.

\bibitem{olfati05}
R.~Olfati-Saber and J.~Shamma, ``Consensus filters for sensor networks and
  distributed sensor fusion,'' in \emph{Proceedings of the 44th IEEE Conference
  on Decision and Control}, 2005, pp. 6698--6703.

\bibitem{fearnhead19}
P.~Fearnhead and G.~Rigaill, ``{Changepoint detection in the presence of
  outliers},'' \emph{{Journal of the American Statistical Association}}, vol.
  114, no. 525, pp. 169--183, 2019.

\end{thebibliography}
\end{document}